\documentclass{article}

\usepackage{subcaption}
\usepackage{amsmath}
\usepackage{array}
\usepackage{arxiv}

\usepackage[utf8]{inputenc}
\usepackage[T1]{fontenc}
\usepackage{hyperref}
\usepackage{url}
\usepackage{booktabs}
\usepackage{amsfonts}
\usepackage{nicefrac}
\usepackage{microtype}
\usepackage{graphicx}
\usepackage{natbib}
\usepackage{doi}

\title{Comparative Analysis of PI and PID Controllers for Level and Flow Control in Coupled Tank Systems}

\author{ \href{https://orcid.org/0009-0003-1388-6752}{\includegraphics[scale=0.06]{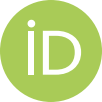}\hspace{1mm}Divyam R. Kakani} \\
	Department of Electronics Engineering\\
	K. J. Somaiya College of Engineering\\
	Mumbai, Maharashtra 400077 \\
	\texttt{divyam.kakani@somaiya.edu} \\
	\And
	\href{https://orcid.org/0009-0001-6452-0895}{\includegraphics[scale=0.06]{orcid.png}\hspace{1mm}Dineshkumar Subendran} \\
	Department of Electronics Engineering\\
	K. J. Somaiya College of Engineering\\
	Mumbai, Maharashtra 400077 \\
	\texttt{d.subendran@somaiya.edu} \\
    \And
	\href{https://orcid.org/0009-0006-0696-577X}{\includegraphics[scale=0.06]{orcid.png}\hspace{1mm}Shivam J. Singh} \\
	Department of Electronics Engineering\\
	K. J. Somaiya College of Engineering\\
	Mumbai, Maharashtra 400077 \\
	\texttt{shivam.js@somaiya.edu} \\
}

\renewcommand{\headeright}{Technical Report}
\renewcommand{\undertitle}{Technical Report}
\renewcommand{\shorttitle}{}

\begin{document}
\maketitle

\begin{abstract}
The comparative study of Proportional-Integral (PI) and Proportional-Integral-Derivative (PID) controllers applied to level and flow control in coupled tank systems is presented in this research work. The coupled tank system, characterized by its nonlinear behavior, was selected due to its relevance in chemical processing industries where precision in liquid level control is crucial. The study evaluates the performance of both controllers under varying conditions, focusing on their ability to handle disturbances and maintain stability. Through experimental data and graphical analysis, it was observed that PID controllers, with their derivative action, provide faster response times and higher accuracy but are more sensitive to noise and harder to tune. In contrast, PI controllers, though slower, offer more stability and are easier to configure for systems where precise control is less critical. These findings highlight the trade-offs between the two control strategies, providing insights into their application depending on system requirements.
\end{abstract}


\section{Introduction}
Accurate control of liquid levels and flow rates is vital in various industrial processes, particularly in chemical processing industries where precision can directly impact product quality and safety. Traditionally, Proportional-Integral-Derivative (PID) controllers have been favored due to their flexibility and robust performance in maintaining desired system behavior. Despite their widespread use, tuning PID controllers for nonlinear systems, such as coupled tank systems, can present significant challenges, as these systems are characterized by complex interactions and variable dynamics. Coupled tank systems are commonly used in industry, where the flow of liquid between interconnected tanks must be precisely controlled, making them a prime candidate for studying the effectiveness of different control strategies.

While PI controllers offer simpler implementation and are well-suited for applications where system dynamics are relatively straightforward, their limitations become apparent in scenarios requiring higher precision or faster response times. This study aims to compare the performance of PI and PID controllers in handling level and flow control within a coupled tank system, emphasizing their advantages and limitations in industrial settings. By analyzing parameters such as proportional gain, integral action time, and derivative action, this work provides a deeper understanding of how each controller type behaves under different operational conditions, offering valuable insights into their suitability for specific control tasks.

\section{Literature Survey}

The design and implementation of control strategies for level and flow regulation in coupled tank systems have been the subject of various research efforts. PID controllers have been predominantly used due to their simplicity and effectiveness in maintaining desired set points. In \cite{jaafar2014development}, the development of a PID controller to regulate the water level in coupled tank systems is described, highlighting the system’s response characteristics under different control parameters.

\cite{kumar2013liquid} examined the application of a fractional PID controller for liquid level control, offering an alternative approach that enhances system stability and transient response. Similarly, \cite{toms2014comparison} compared the performance of a traditional PID controller with a sliding mode controller, demonstrating that while both are effective, the sliding mode controller offers improved robustness in the presence of disturbances.

\cite{fellani2015pid} proposed a MATLAB-based PID controller design for liquid level control in two-tank systems, focusing on enhancing the controller’s accuracy and stability. This study aligns with \cite{ali2020iot}'s work, which explores the implementation of an IoT-enabled level control system utilizing a PLC, thereby expanding the controller's capability for remote monitoring and control.

The use of PLCs in water level management has been further explored by \cite{chakraborty2020development}, who developed a PLC-SCADA-based strategy for water storage control in semi-automated plants, demonstrating improvements in operational efficiency and reliability. \cite{yahya2020design} also investigated a PID-based PLC control module for managing liquid levels, supporting the robustness of PLCs in control applications.

In \cite{suharti2021design}, Suharti et al. studied a PLC-based flow control system specifically tailored for chemical engineering applications, reinforcing the role of PLCs in managing complex process parameters. \cite{zhang2018design} presented an automated feeding control system for tank areas using Siemens PLCs, indicating the adaptability of PLC-based PID controllers in various industrial settings.

Moreover, \cite{soe2019pid} analyzed closed-loop PID control using Siemens PLCs, utilizing TIA V13 software, which underscored the advantages of PLC-integrated control systems in automation processes. Additionally, \cite{daniun2017implementation} implemented an autotuning procedure for PID controllers in PLCs, offering insights into optimizing controller parameters for enhanced performance.

The applications of PID controllers in diverse setups were also examined by \cite{rooholahiintroduction}, who discussed the continuous temperature control using Siemens PLCs, indicating PID controllers' flexibility beyond liquid level applications. In addition, \cite{syufrijal2019construction} introduced a constant pressure control design in water distribution systems, leveraging PLC-based PID methods integrated with IoT for real-time monitoring, which aligns with contemporary trends in control automation. Lastly, \cite{gabor2022implementation} demonstrated the implementation of a PID controller using Siemens PLC, showcasing its practical applicability in industrial settings.

\section{Methodology}

\subsection{System Setup and Configuration}
A coupled tank system was utilized, a common nonlinear setup where tank levels are interdependent. The dynamic inflow and outflow behavior between tanks required precise control for maintaining stability and accuracy. Two primary controllers—PI (Proportional-Integral) and PID (Proportional-Integral-Derivative)—were implemented and tested on this system. Each controller was tuned to specific parameters for optimizing performance in response to disturbances and setpoint changes.

\begin{figure}[htbp]
    \centering
    \includegraphics[width=\linewidth]{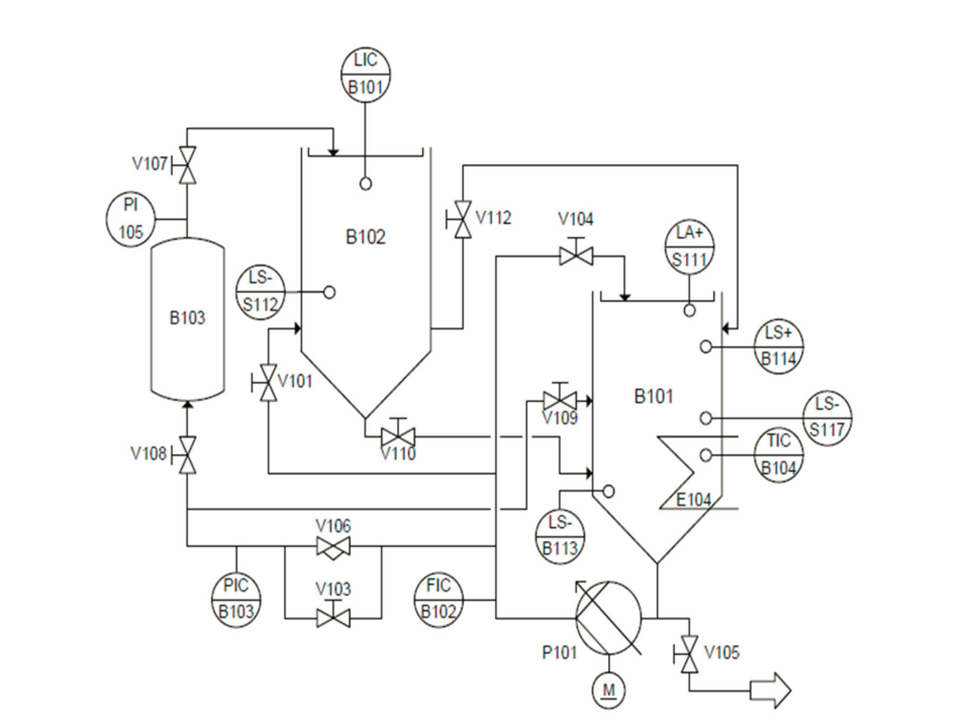}
    \caption{Piping and Instrumentation Diagram of MPS PA Compact Workstation}
\end{figure}

\subsection{Parameter Tuning}
The parameters tuned for each controller included proportional gain (Kp), integral action time (Ti), and derivative action time (Td). Additionally, derivative delay coefficients, proportional and derivative action weightings, and sampling times were adjusted as per the requirements of the PI and PID algorithms. Controller parameters were systematically calibrated to analyze their effects on stability, response time, and accuracy in maintaining tank levels and flow rates.

\subsection{Controller Implementation for Level Control}
The first phase of testing involved setting both controllers to regulate the liquid level within the tanks. Both PI and PID controllers were configured independently, and their outputs were monitored in terms of responsiveness and stability. The performance was assessed by measuring overshoot, steady-state error, and response time to reach the setpoint. Key metrics included the system’s sensitivity to parameter changes, the potential for oscillations, and stability during rapid disturbances.

\begin{table}[h]
    \centering
    \renewcommand{\arraystretch}{1.3}
    \setlength{\tabcolsep}{10pt}
    \caption{Comparison of Parameters of PI and PID Controller for Level Control}
    \begin{tabular}{|l|c|c|}
        \hline
        \textbf{Parameters} & \textbf{PI} & \textbf{PID} \\
        \hline
        Proportional Gain & 124.468 & 516.209 \\
        \hline
        Integral Action Time & 7.220 & 1.047 \\
        \hline
        Derivative Action Time & 0.0 & 2.661543E-1 \\
        \hline
        Derivative Delay Coefficient & 0.1 & 0.1 \\
        \hline
        Proportional Action Weighting & 0.8 & 2.514394E-1\\
        \hline
        Derivative Action Weighting & 0.0 & 0.0 \\
        \hline
        Sampling Time of PID Algorithm & 9.99998E-2 & 9.9998E-2\\
        \hline
    \end{tabular}
    \label{table:level_control_comparison}
\end{table}

\begin{figure}[htbp]
    \centering
    \begin{subfigure}{0.48\linewidth}
        \centering
        \includegraphics[width=\linewidth]{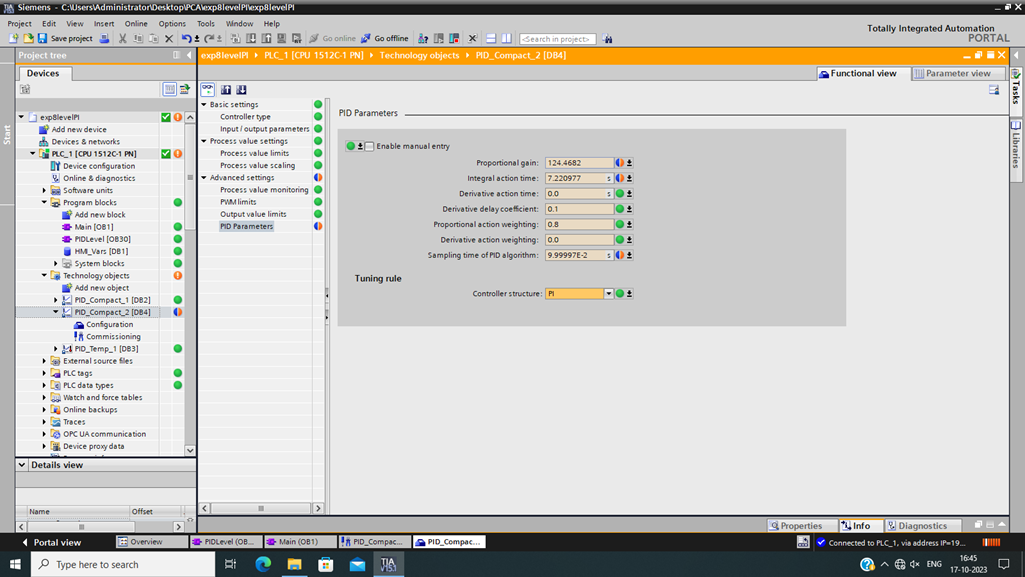}
        \caption{Parameters of Level Control using PI Controller}
        \label{fig:pi_control}
    \end{subfigure}
    \hfill
    \begin{subfigure}{0.48\linewidth}
        \centering
        \includegraphics[width=\linewidth]{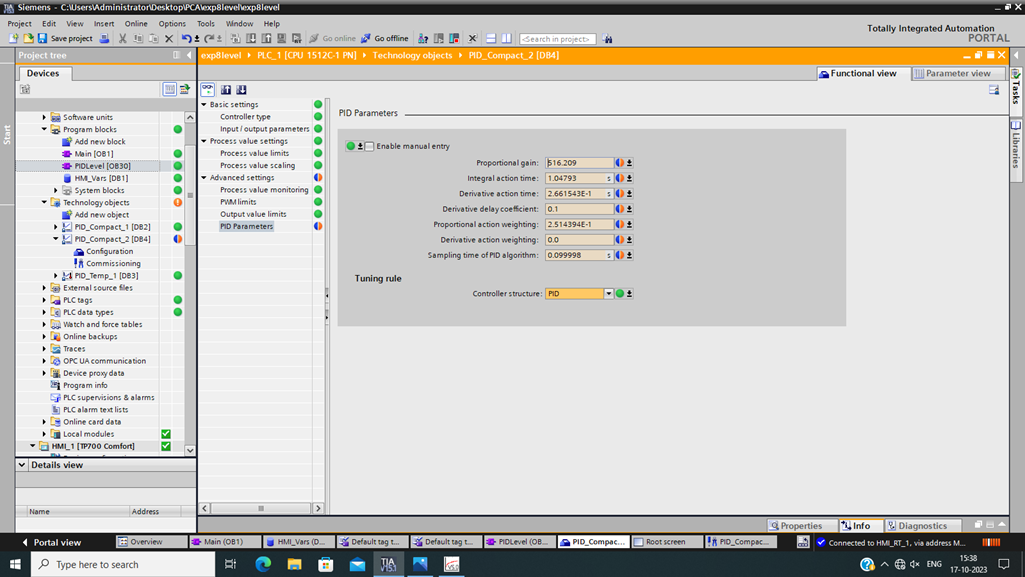}
        \caption{Parameters of Level Control using PID Controller}
        \label{fig:pid_control}
    \end{subfigure}
    \caption{Parameters of PI and PID Controllers for Level Control}
    \label{fig:comparison_level_control}
\end{figure}

\subsection{Controller Implementation for Flow Control using a Pump}
The second phase tested the controllers for flow regulation through a pump. Parameters like proportional gain and integral action time were adjusted to ensure optimal response to desired flow rates. Performance indicators included the system’s ability to reach and maintain the setpoint and handle oscillations, noise, and flow fluctuations. These parameters directly influenced how each controller modulated pump output to maintain desired flow.

\begin{table}[h]
    \centering
    \renewcommand{\arraystretch}{1.3}
    \setlength{\tabcolsep}{10pt}
    \caption{Comparison of Parameters of PI and PID Controller for Flow Control using Pump}
    \begin{tabular}{|l|c|c|}
        \hline
        \textbf{Parameters} & \textbf{PI} & \textbf{PID} \\
        \hline
        Proportional Gain & 6.799 & 1.049 \\
        \hline
        Integral Action Time & 3.174 & 3.688 \\
        \hline
        Derivative Action Time & 0.0 & 4.871 \\
        \hline
        Derivative Delay Coefficient & 0.1 & 0.1 \\
        \hline
        Proportional Action Weighting & 0.8 & 1.0 \\
        \hline
        Derivative Action Weighting & 0.0 & 0.0 \\
        \hline
        Sampling Time of PID Algorithm & 9.99998E-2 & 1.000025E-1 \\
        \hline
    \end{tabular}
    \label{table:flow_control_using_pump_comparison}
\end{table}

\begin{figure}[htbp]
    \centering
    \begin{subfigure}{0.48\linewidth}
        \centering
        \includegraphics[width=\linewidth]{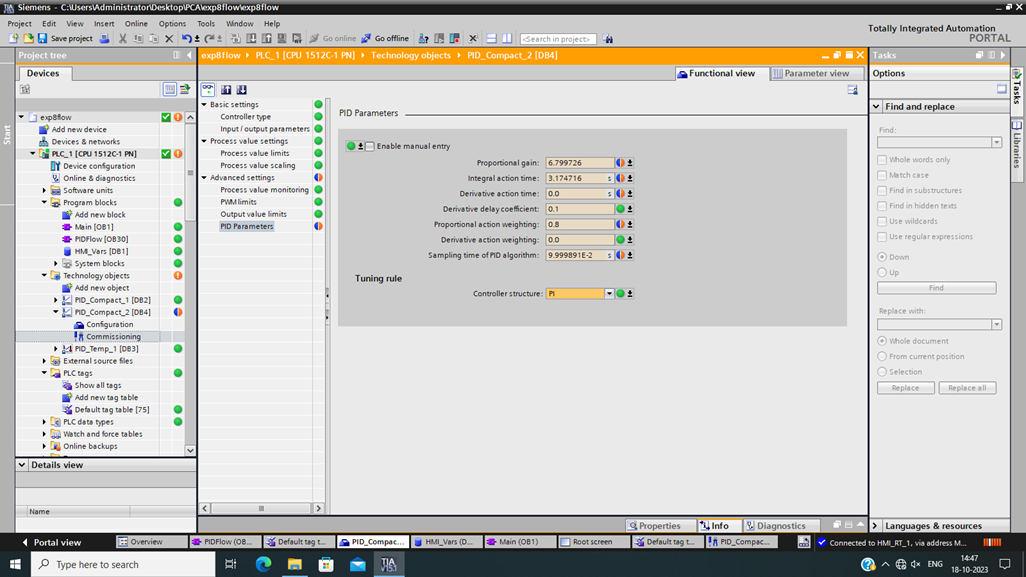}
        \caption{Parameters of Flow Control by Pump using PI Controller}
        \label{fig:pi_control}
    \end{subfigure}
    \hfill
    \begin{subfigure}{0.48\linewidth}
        \centering
        \includegraphics[width=\linewidth]{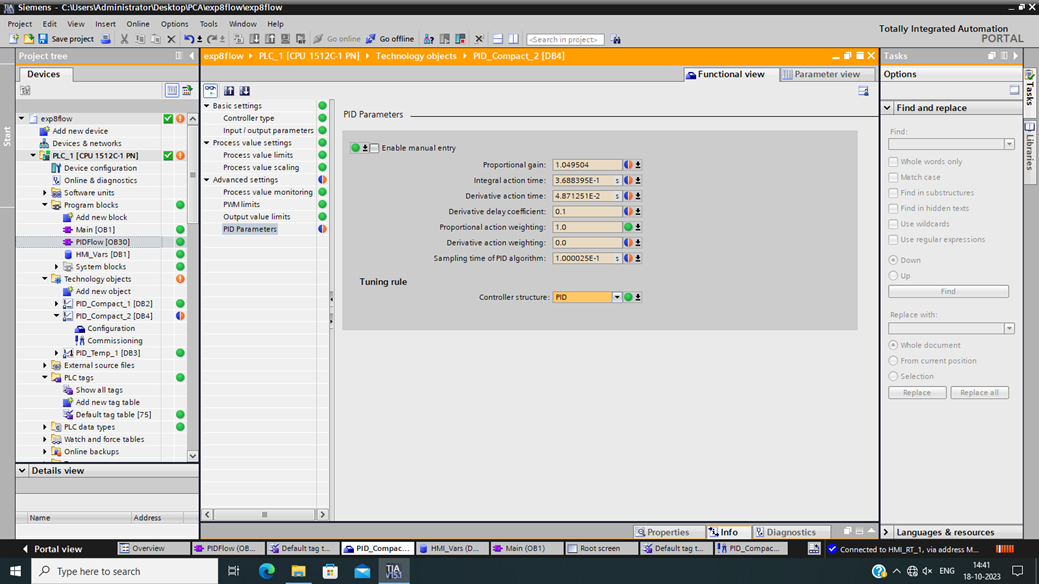}
        \caption{Parameters of Flow Control by Pump using PID Controller}
        \label{fig:pid_control}
    \end{subfigure}
    \caption{Parameters of PI and PID Controllers for Flow Control by Pump}
    \label{fig:comparison_level_control}
\end{figure}

\subsection{Controller Implementation for Flow Control using a Proportional Valve}
Lastly, the controllers were assessed for flow regulation via a proportional valve. This involved observing how each controller managed flow rate setpoints by adjusting the valve opening percentage in response to changes in flow rate and disturbances. Here, the derivative action time and sampling time played significant roles in stability and responsiveness. The controllers were evaluated for their capacity to prevent overshoot, minimize steady-state errors, and handle frequent adjustments.

\begin{table}[h]
    \centering
    \renewcommand{\arraystretch}{1.3}
    \setlength{\tabcolsep}{10pt}
    \caption{Comparison of Parameters of PI and PID Controller for Flow Control using Proportional Valve}
    \begin{tabular}{|l|c|c|}
        \hline
        \textbf{Parameters} & \textbf{PI} & \textbf{PID} \\
        \hline
        Proportional Gain & 15.326 & 6.647 \\
        \hline
        Integral Action Time & 2.489 & 7.981 \\
        \hline
        Derivative Action Time & 0.0 & 1.869 \\
        \hline
        Derivative Delay Coefficient & 0.1 & 0.1 \\
        \hline
        Proportional Action Weighting & 0.8 & 0.971 \\
        \hline
        Derivative Action Weighting & 0.0 & 0.0 \\
        \hline
        Sampling Time of PID Algorithm & 9.99978E-2 & 9.99998E-2 \\
        \hline
    \end{tabular}
    \label{table:flow_control_using_proportional_valve_comparison}
\end{table}

\begin{figure}[htbp]
    \centering
    \begin{subfigure}{0.48\linewidth}
        \centering
        \includegraphics[width=\linewidth]{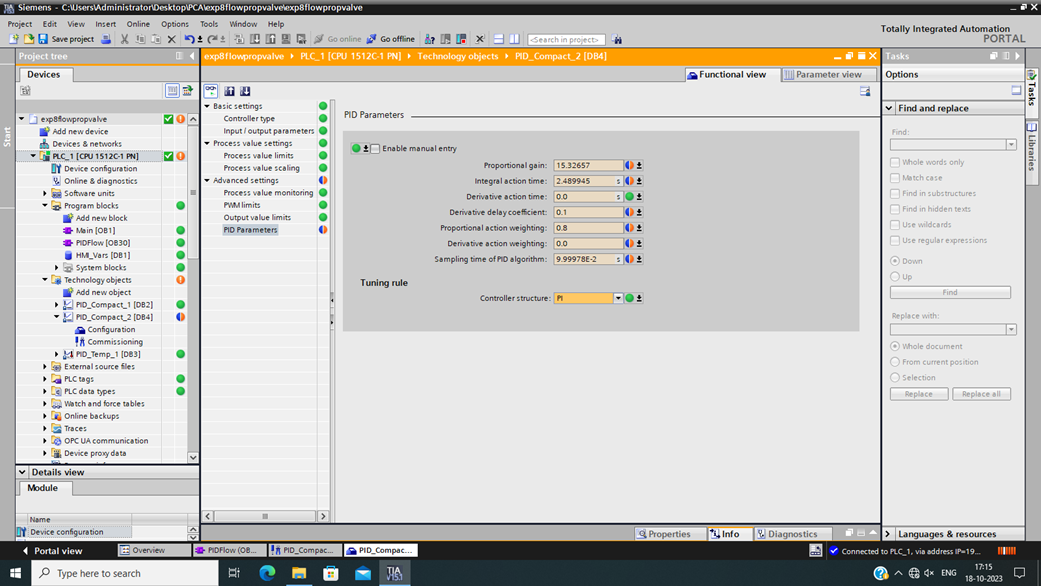}
        \caption{Parameters of Flow Control by Proportional Valve using PI Controller}
        \label{fig:pi_control}
    \end{subfigure}
    \hfill
    \begin{subfigure}{0.48\linewidth}
        \centering
        \includegraphics[width=\linewidth]{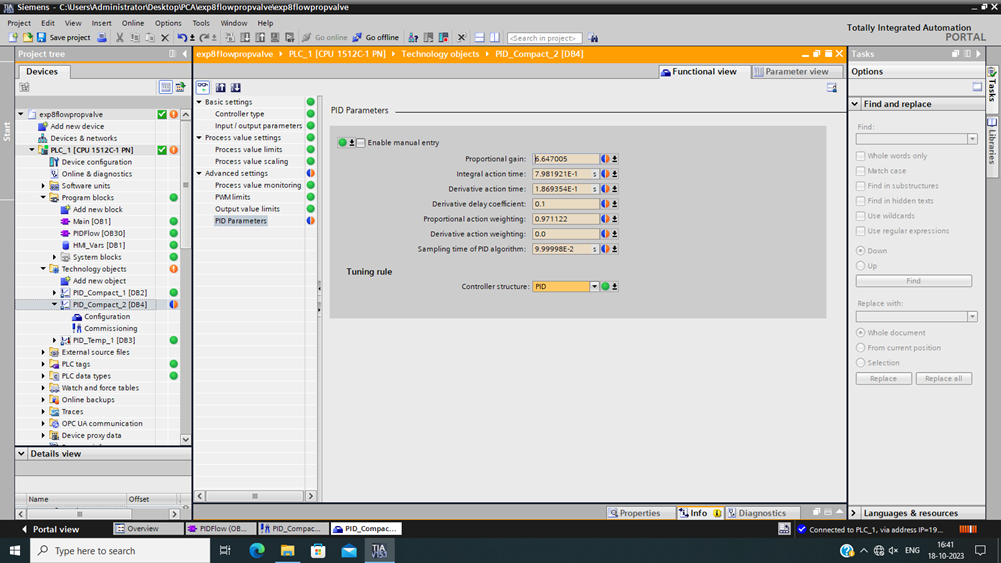}
        \caption{Parameters of Flow Control by Proportional Valve using PID Controller}
        \label{fig:pid_control}
    \end{subfigure}
    \caption{Parameters of PI and PID Controllers for Flow Control by Proportional Valve}
    \label{fig:comparison_level_control}
\end{figure}

\subsection{Data Collection and Analysis}
For each controller and control setup (level and flow control), data was collected on the key performance indicators, including overshoot, response time, and error margin. Comparative graphs and parameter tables were created to illustrate the distinctions in performance between PI and PID controllers. Statistical metrics were computed to quantitatively compare the effectiveness of both controllers under varying conditions and parameter adjustments.

\section{Results and Discussion}

\subsection{Comparative Analysis of PI and PID Controllers for Level Control}

\textbf{1. Proportional Gain:} The proportional gain (Kp) controls the immediate reaction of the controller to the error. The PID controller, with a Kp of 516, responds more aggressively than the PI controller, which has a Kp of 124. This higher gain allows the PID controller to reach setpoints quickly and to address rapid changes in tank level effectively. However, this responsiveness introduces a risk of overshoot and oscillation if the tank level fluctuates rapidly, leading to instability. In contrast, the PI controller’s lower proportional gain results in a slower but smoother response, reducing the risk of overshoot and making it suitable for systems prioritizing stability over response speed.

\textbf{2. Integral Action Time:} The integral action time (Ti) manages the correction of steady-state errors. With a Ti of 1.047 for the PID controller and 7.220 for the PI controller, the PID controller achieves quicker steady-state error correction, suitable for fast response requirements. However, this rapid correction may lead to oscillations, especially under fluctuating conditions. The PI controller, with its longer Ti, provides a gradual adjustment, leading to a more stable level control with minimal overshoot, favoring applications that require sustained stability over rapid error correction.

\textbf{3. Derivative Action Time:} The derivative action time (Td) of 0.266 in the PID controller enhances responsiveness by considering the rate of error change, dampening oscillations and improving response time. This makes the PID controller more adept at managing sudden changes in level, which is particularly useful in preventing overshoot. On the other hand, the PI controller lacks a derivative component, meaning it responds to errors without the added damping effect, which can slow its response time but makes it less sensitive to measurement noise, enhancing stability.

\textbf{4. Derivative Delay Coefficient:} A derivative delay coefficient (N) of 0.1 in both controllers defines the delay in the controller's response to changes in tank level. In the PID controller, this rapid delay coefficient facilitates faster corrective action but increases noise sensitivity, which can exacerbate oscillations under high-gain settings. The PI controller, without a derivative term, is unaffected by this coefficient and remains stable, albeit with a slower response to level changes.

\textbf{5. Proportional Action Weighting:} The proportional action weighting ($\beta$) dictates the influence of the proportional term on the overall control output. The PID controller’s lower weighting of 0.251 makes it less sensitive to minor changes, focusing on significant errors, whereas the PI controller’s $\beta$ of 0.8 provides a more sensitive, albeit slower, reaction to smaller discrepancies. This makes the PID controller better suited for managing abrupt level changes, while the PI controller provides a smoother response, minimizing the risk of overshoot in steady-state scenarios.

\textbf{6. Derivative Action Weighting:} With both controllers having a derivative action weighting ($\alpha$) of 0.0, neither implements pre-emptive adjustments based on error rate predictions. The PID controller, however, still benefits from its base derivative term, which improves oscillation control. The PI controller’s response relies solely on its proportional and integral components, offering stability at the expense of rapid responsiveness to changing tank levels.

\textbf{7. Sampling Time of PID Algorithm:} The sampling time determines how often the controllers adjust outputs. The PID controller’s sampling time of 0.099998 seconds provides a high-frequency update rate, suitable for dynamic control applications requiring precise and fast response. The PI controller, with a sampling time of 0.099997 seconds, updates slightly slower, which reduces sensitivity to noise and improves stability at the cost of a marginally slower response to level changes.

\begin{figure}[htbp]
    \centering
    \begin{subfigure}{0.48\linewidth}
        \centering
        \includegraphics[width=\linewidth]{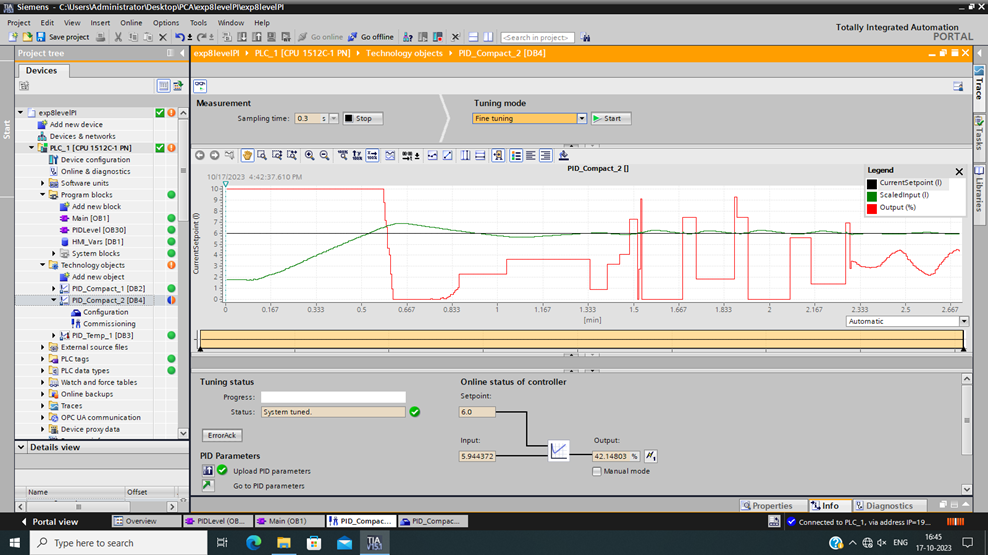}
        \caption{Response of PI Controller for Level Control}
        \label{fig:pi_control}
    \end{subfigure}
    \hfill
    \begin{subfigure}{0.48\linewidth}
        \centering
        \includegraphics[width=\linewidth]{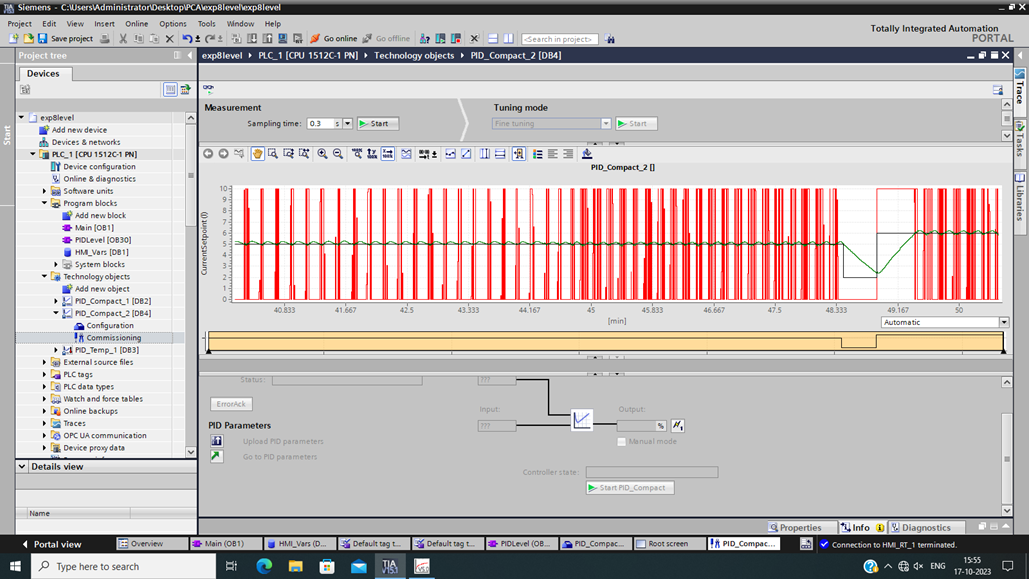}
        \caption{Response of PID Controller for Level Control}
        \label{fig:pid_control}
    \end{subfigure}
    \caption{Comparison of PI and PID Controllers for Level Control}
    \label{fig:comparison_level_control}
\end{figure}

The results indicate that the PID controller, with higher proportional gain and shorter integral action time, is better suited for applications demanding fast and precise level control, particularly where tank levels experience rapid fluctuations. However, this setup also makes the PID controller more sensitive to noise and prone to overshoot. Conversely, the PI controller provides a stable and gradual response, with a lower proportional gain and longer integral action time, making it appropriate for systems where stability is a priority over rapid adjustment. Overall, the PID controller is advantageous in dynamic environments requiring quick error correction, while the PI controller excels in scenarios where stability is more critical than response speed.

\subsection{Comparative Analysis of PI and PID Controllers for Flow Control using Pump}

\textbf{1. Proportional Gain:} The proportional gain (Kp) determines how aggressively the controller responds to errors. The PID controller exhibited a high proportional gain of 516.209 compared to 124.468 for the PI controller, leading to a more aggressive response in level adjustments. This high gain enabled the PID controller to correct level discrepancies quickly, but also introduced a risk of overshoot and oscillation, especially in response to rapid disturbances. The PI controller, with a lower proportional gain, provided a slower but stable response, making it more suitable for conditions prioritizing stability over quick corrections.

\textbf{2. Integral Action Time:} The integral action time (Ti) addresses the accumulation of past errors, helping to eliminate steady-state error over time. With an integral action time of 1.047 for PID and 7.220 for PI, the PID controller corrected steady-state errors more rapidly. However, this also made the system more prone to oscillatory behavior, as the quick integral response can contribute to overcompensation. The PI controller, with its longer Ti, demonstrated smoother performance in maintaining tank levels, reducing the risk of oscillation but at the cost of a slower correction rate.

\textbf{3. Derivative Action Time:} The derivative action time (Td) of 0.266 for the PID controller enabled it to account for the rate of change in level, allowing faster damping of oscillations and an improved response time. This derivative term made the PID controller more responsive to sudden level fluctuations, which is crucial for minimizing overshoot. The PI controller, lacking a derivative term (Td = 0), responded more gradually to changes, making it less susceptible to noise amplification but slower in handling sudden level adjustments.

\textbf{4. Derivative Delay Coefficient:} Both controllers were assigned a derivative delay coefficient (N) of 0.1, representing a quick response to error signals. For the PID controller, this minimized response time to disturbances, but also increased the sensitivity to noise, as rapid corrective actions could amplify minor fluctuations. The PI controller, unaffected by derivative action, demonstrated a stable but slower adaptation to disturbances.

\textbf{5. Proportional Action Weighting:} The proportional action weighting ($\beta$) determines how much the proportional term influences the output. The PID controller’s weighting of 0.251 contrasted with the PI controller’s 0.8, making the PID controller less sensitive to small changes in error while still effective at larger error corrections. This weighting balance in the PID controller contributed to its ability to handle rapid disturbances, though it required careful tuning to prevent instability.

\textbf{6. Derivative Action Weighting:} With a derivative action weighting ($\alpha$) of 0 for both controllers, neither emphasized pre-emptive corrections based on predicted error behavior, resulting in responses that mainly addressed current and past errors. The absence of a derivative weighting impact for the PI controller allowed a steady output unaffected by future predictions, while the PID controller’s base derivative term still helped control oscillations effectively.

\textbf{7. Sampling Time of PID Algorithm:} Sampling time affects how often the controllers process the error and adjust output. The PID controller’s sampling time of 0.099998 seconds offered a higher frequency of adjustments, allowing faster responses to dynamic changes. In contrast, the PI controller’s sampling time of 0.099997 seconds provided a marginally slower update rate, contributing to stability but at the cost of slower response to rapid level changes.

\begin{figure}[htbp]
    \centering
    \begin{subfigure}{0.48\linewidth}
        \centering
        \includegraphics[width=\linewidth]{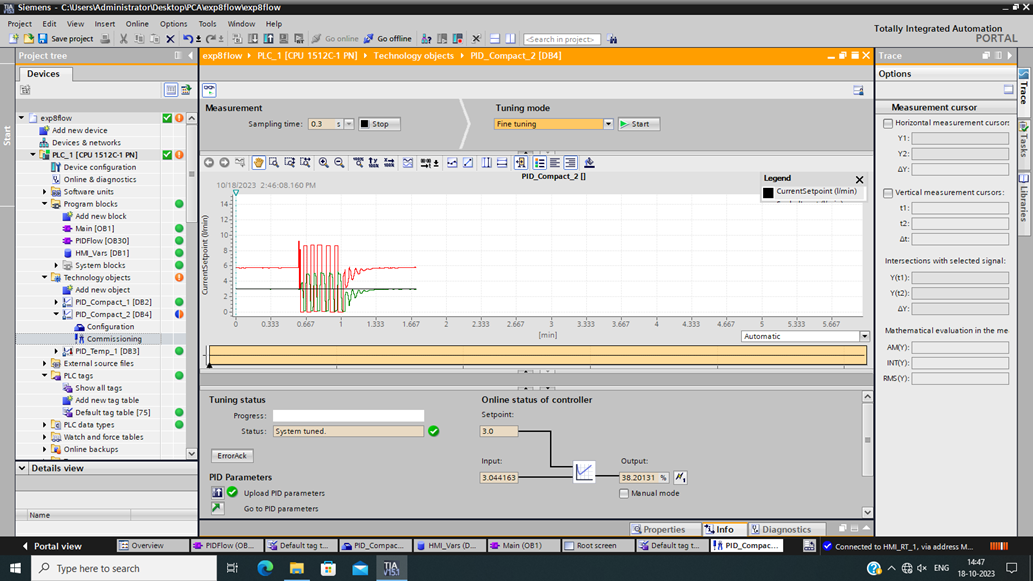}
        \caption{Response of PI Controller for Flow Control using Pump}
        \label{fig:pi_control}
    \end{subfigure}
    \hfill
    \begin{subfigure}{0.48\linewidth}
        \centering
        \includegraphics[width=\linewidth]{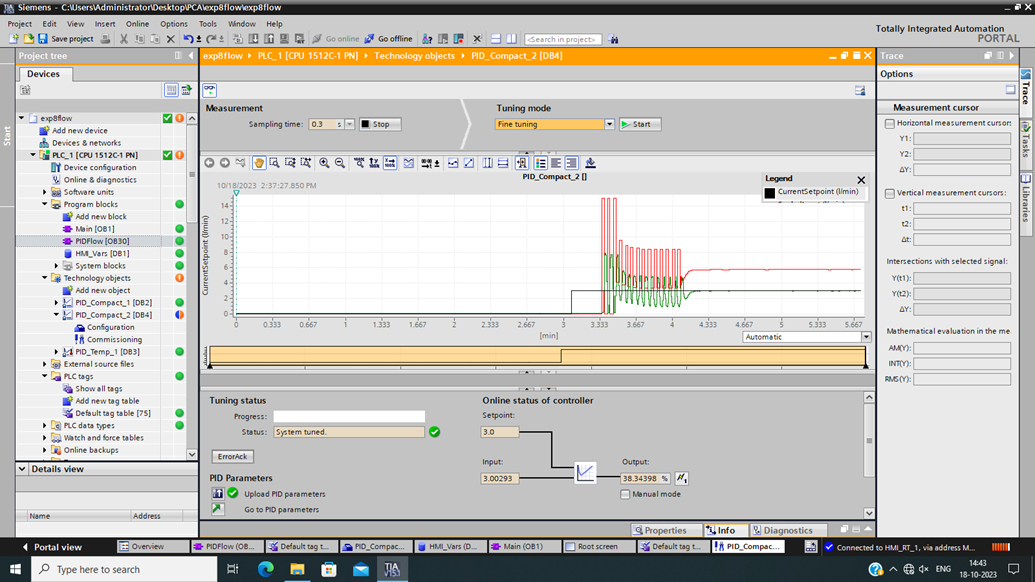}
        \caption{Response of PID Controller for Flow Control using Pump}
        \label{fig:pid_control}
    \end{subfigure}
    \caption{Comparison of PI and PID Controllers for Flow Control using Pump}
    \label{fig:comparison_level_control}
\end{figure}

Overall, the PID controller’s higher proportional gain and shorter integral action time allowed for quicker, more precise level adjustments, but introduced susceptibility to noise and overshoot. The PI controller, with lower proportional gain and longer integral action time, provided a smoother response, making it more stable for applications where response speed is less critical. The PID controller excelled in dynamic environments requiring rapid corrections, while the PI controller offered robustness in systems where stability is prioritized over fast response.

\subsection{Comparative Analysis of PI and PID Controllers for Flow Control using Proportional Valve}

\textbf{1. Proportional Gain:} The proportional gain (Kp) determines the initial intensity of the controller’s response to errors in flow rate. The PID controller, with a higher Kp, causes the valve to open or close more rapidly to correct flow discrepancies, which is beneficial in applications requiring precise flow control. However, this high responsiveness can lead to instability and oscillations if the system is not properly tuned. In contrast, the PI controller’s lower Kp promotes stability, with the integral term compensating for any steady-state errors, making it suitable for maintaining a consistent flow rate despite disturbances.

\textbf{2. Integral Action Time:} The integral action time (Ti) manages how quickly steady-state errors are corrected. A higher Ti value in the PI controller allows it to maintain a stable flow rate, as it responds more gradually to fluctuations. In applications where the flow rate needs to be maintained consistently, the PI controller’s higher Ti provides smoother control and stability. For scenarios demanding faster adjustments, the PID controller’s lower Ti corrects steady-state errors more quickly but requires careful tuning to prevent oscillatory behavior.

\textbf{3. Derivative Action Time:} The derivative action time (Td) of the PID controller provides responsiveness to the rate of change in flow, with a Td of 1.869 enabling it to respond to rapid setpoint changes. However, this responsiveness makes it slower at managing steady-state disturbances in flow control. The PI controller, lacking a derivative term, avoids the complexities of tuning the derivative parameter and provides faster, stable flow adjustments. This feature makes PI more effective for flow control in systems with lower oscillation tolerance.

\textbf{4. Derivative Delay Coefficient:} The derivative delay coefficient (N) of 0.1 indicates a slight lag before the derivative response takes effect. For the PID controller, this delay in the derivative action can cause instability and oscillations in the flow control system if not managed properly, as the delayed response may overcompensate. The PI controller is unaffected by this delay, maintaining a stable output that is less sensitive to noise, suitable for steady-state applications with minimal fluctuation.

\textbf{5. Proportional Action Weighting:} Proportional action weighting ($\beta$) affects the controller’s responsiveness to flow rate changes. The PI controller’s weighting of 0.8 produces a faster response to error changes but may lead to minor oscillations, making it effective for systems where quick response to flow adjustments is needed. The PID controller, with a lower $\beta$ of 0.971, responds less aggressively to minor errors, contributing to a more stable output in applications with sensitive flow dynamics, though it may require more precise tuning to prevent overshoot.

\textbf{6. Derivative Action Weighting:} The derivative action weighting ($\alpha$) of 0.0 in both controllers limits the influence of the derivative term on the output. This omission benefits the PI controller, which avoids the noise sensitivity associated with derivative action, making it suitable for maintaining steady flow rates with minimal oscillation. The PID controller, although having a derivative term, maintains stability in flow rate adjustments, allowing for finer tuning based on specific flow control needs.

\textbf{7. Sampling Time of PID Algorithm:} The sampling time determines the update rate of the controller. A shorter sampling time in the PI controller improves its responsiveness to changes in flow rate, aiding in maintaining stability despite minor disturbances. In applications requiring precise flow adjustments, the PID controller’s longer sampling time of 1.35 seconds ensures stability but requires careful adjustment to balance between responsiveness and noise sensitivity.

\begin{figure}[htbp]
    \centering
    \begin{subfigure}{0.48\linewidth}
        \centering
        \includegraphics[width=\linewidth]{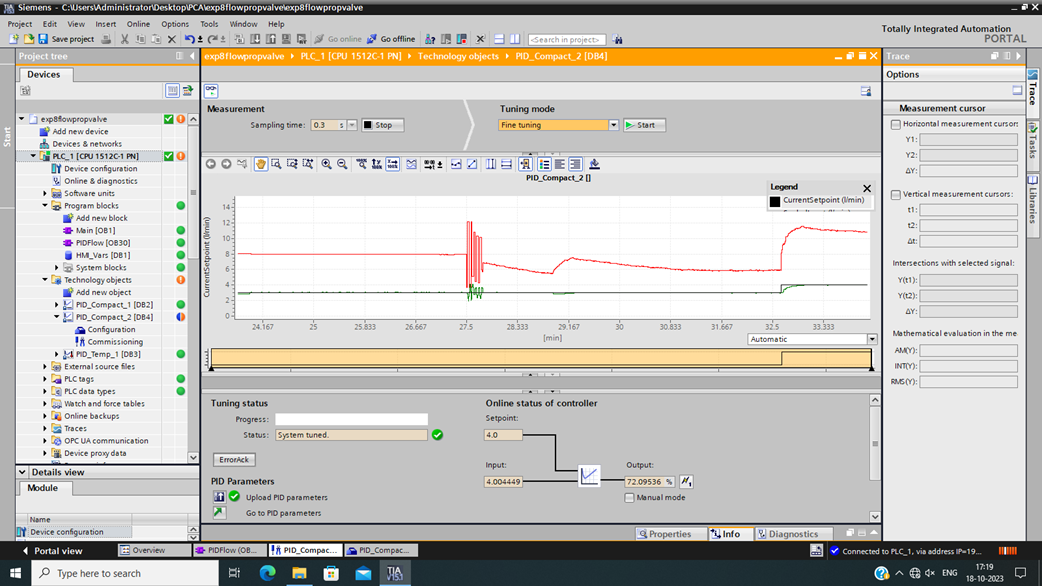}
        \caption{Response of PI Controller for Flow Control using Proportional Valve}
        \label{fig:pi_control}
    \end{subfigure}
    \hfill
    \begin{subfigure}{0.48\linewidth}
        \centering
        \includegraphics[width=\linewidth]{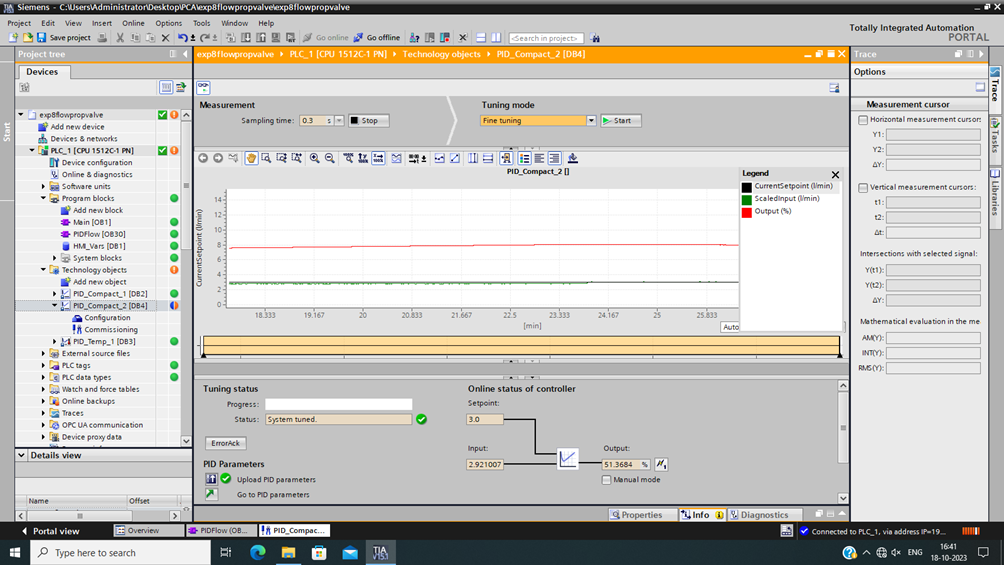}
        \caption{Response of PID Controller for Flow Control using Proportional Valve}
        \label{fig:pid_control}
    \end{subfigure}
    \caption{Comparison of PI and PID Controllers for Flow Control using Proportional Valve}
    \label{fig:comparison_level_control}
\end{figure}

The PID controller’s higher proportional gain and shorter integral action time are advantageous for applications demanding rapid and precise flow control but require careful tuning to prevent instability. The PI controller provides a slower, stable response due to its higher integral action time and lack of a derivative term, making it ideal for applications prioritizing steady-state control and stability. The PI controller excels in stable environments, whereas the PID controller is preferred in dynamic systems needing fast correction with minimal overshoot.

\subsection{Discussion}

For \textbf{level control}, the PID controller demonstrated a faster and more aggressive response due to its higher proportional gain (Kp) and shorter integral action time (Ti). This configuration allowed the PID controller to handle rapid level fluctuations effectively, making it suitable for dynamic applications requiring quick adjustments. However, this high responsiveness also led to increased noise sensitivity and the potential for overshoot and oscillations, particularly in systems with frequent disturbances. Conversely, the PI controller, with a lower Kp and longer Ti, offered a slower, more gradual response. This setup improved system stability, making the PI controller ideal for steady-state applications where stability is prioritized over rapid response.

In \textbf{flow control} scenarios, the performance of each controller varied based on the control component used, whether a pump or a proportional valve. The PID controller excelled in applications needing precise flow adjustments and quick response to setpoint changes. However, the derivative action (Td) introduced additional noise sensitivity and required careful tuning to mitigate oscillations, especially when using a proportional valve. The PI controller, with a longer Ti and the absence of a derivative term, provided a smoother flow control response. This setup minimized noise amplification, which is advantageous in applications where consistent flow is crucial. The PI controller’s lower proportional action weighting ($\beta$) further contributed to system stability, allowing for a less aggressive response to flow rate errors, which is beneficial in systems prone to steady-state operation.

Across both level and flow control applications, the study underscores that the PID controller is better suited for environments requiring rapid and precise adjustments, albeit with an increased need for tuning to manage noise and instability. In contrast, the PI controller excels in systems prioritizing stability, offering a robust and smooth control response at the expense of response speed. The choice between PI and PID controllers, therefore, depends on the specific control requirements and desired trade-offs between response speed and system stability.

\section{Future Work}

Future research could explore more advanced control strategies like Model Predictive Control (MPC) or adaptive control, which may better handle nonlinear systems like the coupled tank setup. Integrating machine learning could also enhance automatic tuning of controllers, reducing the need for manual adjustments and improving system performance over time. Noise filtering techniques, such as Kalman filters, would help reduce the sensitivity of PID controllers to measurement noise, addressing one of their key drawbacks.

Extending the study to multi-input-multi-output (MIMO) systems could provide insights into how PI and PID controllers perform in more complex industrial setups. Additionally, implementing these controllers in real-time distributed control systems (DCS) would offer practical insights into their scalability and robustness in managing large-scale processes. Research into hybrid control strategies that combine PID with nonlinear compensators could further improve system performance, especially in dynamic or unstable environments. Finally, testing these controllers in other industrial contexts, such as temperature or pressure control, could expand their applicability and optimize their tuning for a range of process controls.

\section{Conclusion}

This study compared PI and PID controllers for level and flow control in a nonlinear coupled tank system. The PID controller, with its higher proportional gain and derivative action, provided faster response times and reduced steady-state errors, making it suitable for applications requiring quick and precise control. However, it was more sensitive to noise and required careful tuning to prevent instability. In contrast, the PI controller, though slower in response, offered greater stability and simplicity, making it preferable for systems where robustness and ease of implementation are more important than speed.

In level control applications, the PID controller demonstrated superior performance in terms of settling time and accuracy, but at the cost of increased noise sensitivity. The PI controller, while slower, provided smoother and more stable control. In flow control scenarios, the PID controller excelled in dynamic response, especially with the pump, but required fine-tuning to avoid overshoot. The PI controller performed better with the proportional valve, offering stable and gradual control with minimal oscillations.

Overall, the choice between PI and PID controllers depends on the specific process requirements: the PID controller is better suited for fast, precise control, while the PI controller is ideal for systems prioritizing stability and ease of tuning.

\bibliographystyle{plainnat}
\setcitestyle{numbers}

\begin{thebibliography}{14}
\providecommand{\natexlab}[1]{#1}
\providecommand{\url}[1]{\texttt{#1}}
\expandafter\ifx\csname urlstyle\endcsname\relax
  \providecommand{\doi}[1]{doi: #1}\else
  \providecommand{\doi}{doi: \begingroup \urlstyle{rm}\Url}\fi

\bibitem[Ali et~al.(2020)Ali, Miry, and Salman]{ali2020iot}
Methaq~A Ali, Abbas~Hussein Miry, and Tariq~M Salman.
\newblock Iot based water tank level control system using plc.
\newblock In \emph{2020 International Conference on Computer Science and Software Engineering (CSASE)}, pages 7--12. IEEE, 2020.

\bibitem[Chakraborty et~al.(2020)Chakraborty, Choudhury, Das, and Paul]{chakraborty2020development}
K~Chakraborty, MG~Choudhury, S~Das, and S~Paul.
\newblock Development of plc-scada based control strategy for water storage in a tank for a semi-automated plant.
\newblock \emph{Journal of Instrumentation}, 15\penalty0 (04):\penalty0 T04007, 2020.

\bibitem[Daniun et~al.(2017)Daniun, Awtoniuk, and Sa{\l}at]{daniun2017implementation}
Marcin Daniun, Micha{\l} Awtoniuk, and Robert Sa{\l}at.
\newblock Implementation of pid autotuning procedure in plc controller.
\newblock In \emph{ITM Web of Conferences}, volume~15, page 05009. EDP Sciences, 2017.

\bibitem[Fellani and Gabaj(2015)]{fellani2015pid}
Mostafa~Abdul Fellani and Aboubaker~M Gabaj.
\newblock Pid controller design for two tanks liquid level control system using matlab.
\newblock \emph{International Journal of Electrical and Computer Engineering}, 5\penalty0 (3):\penalty0 436, 2015.

\bibitem[Gabor and Livint(2022)]{gabor2022implementation}
Georgel Gabor and Gheorghe Livint.
\newblock Implementation of a pid controller using siemens plc.
\newblock In \emph{2022 International Conference and Exposition on Electrical And Power Engineering (EPE)}, pages 593--596. IEEE, 2022.

\bibitem[Jaafar et~al.(2014)Jaafar, Hussien, Selamat, Aras, and Rashid]{jaafar2014development}
Hazriq~Izzuan Jaafar, SYS Hussien, NA~Selamat, MSM Aras, and MZA Rashid.
\newblock Development of pid controller for controlling desired level of coupled tank system.
\newblock \emph{International Journal of Innovative Technology and Exploring Engineering}, 3\penalty0 (9):\penalty0 32--36, 2014.

\bibitem[Kumar et~al.(2013)Kumar, Vashishth, and Rai]{kumar2013liquid}
Arun Kumar, Munish Vashishth, and Lalit Rai.
\newblock Liquid level control of coupled tank system using fractional pid controller.
\newblock \emph{International Journal of Emerging Trends in Electrical and Electronics}, 3\penalty0 (1):\penalty0 61--64, 2013.

\bibitem[Rooholahi et~al.()Rooholahi, Roohollahi, and Galyautdinova]{rooholahiintroduction}
Babak Rooholahi, M~Nabi Roohollahi, and Alina Galyautdinova.
\newblock Introduction and implementation of continuous temperature pid controller in siemens plcs.

\bibitem[Soe et~al.(2019)Soe, San, et~al.]{soe2019pid}
Yin~Yin Soe, Pann~Ei San, et~al.
\newblock Pid closed-loop control analysis for automation with siemens plc using tia v13.
\newblock \emph{International Journal Of All Research Writings}, 2\penalty0 (3):\penalty0 200--208, 2019.

\bibitem[Suharti et~al.(2021)Suharti, Kristiana, and Amalia]{suharti2021design}
PH~Suharti, HM~Kristiana, and RN~Amalia.
\newblock Design of flow control system based on plc armfield pressure control module in chemical engineering laboratory.
\newblock In \emph{IOP Conference Series: Materials Science and Engineering}, volume 1073, page 012053. IOP Publishing, 2021.

\bibitem[Syufrijal et~al.(2019)Syufrijal, Rif’an, et~al.]{syufrijal2019construction}
Syufrijal Syufrijal, M~Rif’an, et~al.
\newblock Construction design system of constant pressure control in water distribution system with pid method using plc based on iot.
\newblock In \emph{Journal of Physics: Conference Series}, volume 1402, page 022060. IOP Publishing, 2019.

\bibitem[Toms and Hepsiba(2014)]{toms2014comparison}
Teena Toms and D~Hepsiba.
\newblock Comparison of pid controller with a sliding mode controller for a coupled tank system.
\newblock \emph{International Journal of Engineering Research \& Technology}, 3\penalty0 (2):\penalty0 151--154, 2014.

\bibitem[Yahya et~al.(2020)Yahya, Jadmiko, Wijayanto, and Tahtawi]{yahya2020design}
S~Yahya, SW~Jadmiko, K~Wijayanto, and ARA Tahtawi.
\newblock Design and implementation of training module for control liquid level on tank using pid method based plc.
\newblock In \emph{IOP Conference Series: Materials Science and Engineering}, volume 830, page 032065. IOP Publishing, 2020.

\bibitem[Zhang and Sun(2018)]{zhang2018design}
Wei Zhang and Shuai Sun.
\newblock Design of automatic feeding control system in tank area based on siemens plc.
\newblock In \emph{2018 Chinese Control And Decision Conference (CCDC)}, pages 3622--3627. IEEE, 2018.

\end{thebibliography}

\end{document}